\documentclass[%
reprint,
amsmath,amssymb,
aps,
pra,
]{revtex4-2}
\usepackage{float}
\usepackage{graphicx}
\usepackage{dcolumn}
\usepackage{bm}

\begin{document}

\preprint{APS/123-QED}

\title{All-optical generation of deterministic squeezed Schr\"odinger-cat states}

\author{Zhucheng Zhang,$^{1}$ Lei Shao,$^1$ Wangjun Lu,$^{1,2}$  and Xiaoguang Wang$^{1,3}$}
\email{xgwang1208@zju.edu.cn}
\affiliation{$^{1}$Zhejiang Institute of Modern Physics, School of Physics, Zhejiang University, Hangzhou 310027, China\\
$^{2}$Department of Maths and Physics, Hunan Institute of Engineering, Xiangtan 411104, China\\
$^{3}$Graduate School of China Academy of Engineering Physics, Beijing 100193, China}
\date{\today}

\begin{abstract}
Quantum states are important resources and their preparations are essential prerequisites to all quantum technologies. However, they are extremely fragile due to the inevitable dissipations. Here, an all-optical generation of a deterministic squeezed Schr$\ddot{\mathrm{o}}$dinger-cat state based on dissipation is proposed. Our system is based on the Fredkin-type interaction between three optical modes, one of which is subject to coherent two-photon driving and the rest are coherent driving. We show that an effective degenerate three-wave mixing process can be engineered in our system, which can cause the simultaneous loss of two photons, resulting in the generation of a deterministic squeezed Schr$\ddot{\mathrm{o}}$dinger-cat state. More importantly, by controlling the driving fields in our system, the two-photon loss can be adjustable, which can accelerate the generation of squeezed Schr$\ddot{\mathrm{o}}$dinger-cat states.
Besides, we exploit the squeezed Schr$\ddot{\mathrm{o}}$dinger-cat states to estimate the phase in the optical interferometer, and show
that the quantum Fisher information about the phase can reach the Heisenberg limit in the limit of a large photon number. Meanwhile, it can have an order of magnitude factor improvement over the Heisenberg limit in the low-photon-number regime, which is very valuable for fragile systems that cannot withstand large photon fluxes. This work proposes an all-optical scheme to deterministically prepare the squeezed Schr$\ddot{\mathrm{o}}$dinger-cat state with high speed and can also be generalized to other physical platforms.
\end{abstract}

\maketitle

\section{Introduction}\label{section1}
Schr$\ddot{\mathrm{o}}$dinger cat originally referred to a cat in a superposition of being dead versus alive, which is first introduced by Schr$\ddot{\mathrm{o}}$dinger to question the Copenhagen interpretation of quantum mechanics \cite{Sch1935}. The cat being alive is macroscopically
distinguishable from the cat being dead. In deference to Schr$\ddot{\mathrm{o}}$dinger's paradox of the cat, the superposition of the two coherent states with large and the same amplitude but a phase shift of $\pi$ is called the Schr$\ddot{\mathrm{o}}$dinger-cat state \cite{MOS1997,BCS2012}. According to the difference of the photon number distributions, the Schr$\ddot{\mathrm{o}}$dinger-cat state can generally be divided into three cases \cite{MOS1997,BCS2012}: even coherent state (ECS) with even number distribution, odd coherent state (OCS) with odd number distribution, and Yurke-Stoler coherent state (YSCS) with a Poisson number distribution. Since the YSCS has a Poisson number distribution, it can be generated by a unitary time evolution with a Kerr-type nonlinearity \cite{BCS2012,SM1997,SB1997,JQ2016}. The ECS and the OCS, however, can not be prepared by a unitary time evolution. To generate the ECS and the OCS, one can exploit the interaction with other systems, such as atom, followed by selective quantum measurements \cite{BCS2012,CC1997,LD1996,SD2008,WG2015,YX2020,YH2021}. Besides, in stark contrast to dynamically transient preparation of the YSCS, the stable ECS and OCS can be deterministically generated by engineering a two-photon loss \cite{CC1993,LG1994,RI2008,HT2013,MM2014,MJ2014,ZL2015,ST2018,RL2020,AG2020,WQ2021}, which is very important for realistic applications. Schr$\ddot{\mathrm{o}}$dinger-cat states are not only fundamental studies of quantum mechanics, but also lead to a boom in quantum information science.

Schr$\ddot{\mathrm{o}}$dinger-cat states have moved from basic research to actual quantum technologies, such as quantum computation \cite{TC2003,JG2014,NO2016,WC2021} and quantum metrology \cite{MK2011,JC2014,LP2018}, etc. It has been shown that Schr$\ddot{\mathrm{o}}$dinger-cat states have remarkable applications in error correction codes \cite{BM2015,FG2008,DA2013,YH2021PRR} and geometric quantum computation \cite{YH2022}. For quantum metrology based on optical interferometer, path-entangled Schr$\ddot{\mathrm{o}}$dinger-cat states can reach the estimation precision with Heisenberg limit (HL) for the linear phase and super-Heisenberg limit for the nonlinear phase in the limit of the large photon number \cite{JC2014}. Unfortunately, for the phase estimation in the optical interferometer, it has been proved that the estimation precision depends mainly on the Mandel $\mathcal{Q}$ parameter of the quantum states, and the correlations between the paths (i.e., path-entangled quantum states) can contribute at most a factor of 2 enhancement \cite{JS2015}. Meanwhile, in a practical situation, we must consider the capacity of the sample to withstand the photon fluxes, that is, we should focus on the measurements of the fragile systems in the low-photon-number regime.

To improve the estimation precision in the low-photon-number regime, many quantum states (including N00N state \cite{HL2002}, squeezed-entangled state \cite{PA2016},
entangled even squeezed state \cite{LS2020}, etc.) have been considered. However, the particularly promising approach is to squeeze a non-Gaussian state, and the squeezed Schr$\ddot{\mathrm{o}}$dinger-cat state is the one that has a large Mandel $\mathcal{Q}$ parameter, which can have a threefold improvement over the optimal Gaussian state \cite{PA2016}. Besides, the squeezed Schr$\ddot{\mathrm{o}}$dinger-cat states also have been shown to have a significant role in quantum error correction \cite{DS2022}. For quantum technologies with squeezed Schr$\ddot{\mathrm{o}}$dinger-cat states, the preparation fidelity is crucial for realizing quantum advantage.  There are some schemes to prepare the squeezed Schr$\ddot{\mathrm{o}}$dinger-cat states \cite{AO2007,JE2015,KH2015}, but most of them rely on additional quantum measurements. Meanwhile, the preparation fidelities are only over 60$\%$ in experiments. To our knowledge, there is still a lack of research to deterministically prepare the squeezed Schr$\ddot{\mathrm{o}}$dinger-cat states.

In this paper, we propose an all-optical scheme to deterministically prepare the squeezed Schr$\ddot{\mathrm{o}}$dinger-cat states. In our scheme,
based on the Fredkin-type interaction \cite{GJ1989,RB2016,YY2019} between three optical modes, an effective degenerate three-wave mixing process is engineered, which results
in the generation of the two-photon loss in our system. Then the squeezed Schr$\ddot{\mathrm{o}}$dinger-cat states can be generated deterministically. The sizes of the squeezed Schr$\ddot{\mathrm{o}}$dinger-cat states and their degrees of squeezing are well controllable by adjusting the driving fields of the optical modes. Moreover, the generation time of the squeezed Schr$\ddot{\mathrm{o}}$dinger-cat states can also be greatly shortened. With the generated squeezed Schr$\ddot{\mathrm{o}}$dinger-cat states in our scheme, we also estimate the linear phase in the optical interferometer. We find that the quantum Fisher information (QFI) about the phase can reach the HL in the limit of the large photon number, meanwhile, it can have an order of magnitude factor enhancement to the HL in the low-photon-number regime, which is valuable for the fragile systems to be measured.

This paper is organized as follows. In Sec.~\ref{sec:2}, we introduce the system model and discuss experimental feasibility, then analyze the dynamic evolution. In Sec.~\ref{sec:3}, by adjusting the driving fields of the optical modes in our system, we obtain squeezed Schr$\ddot{\mathrm{o}}$dinger-cat states with different sizes and squeezing degrees, and analyze the robustness of the generated squeezed Schr$\ddot{\mathrm{o}}$dinger-cat states to the system loss. Besides, in Sec.~\ref{sec:4}, we exploit the generated squeezed Schr$\ddot{\mathrm{o}}$dinger-cat states to estimate the phase of the optical interferometer, and compare the OFI about phase with the perfect squeezed Schr$\ddot{\mathrm{o}}$dinger-cat states. Finally, we summarize our conclusions in Sec.~\ref{sec:5}.

\section{Model and Hamiltonian}\label{sec:2}

\begin{figure}[t]
    \centering\includegraphics[scale=0.55]{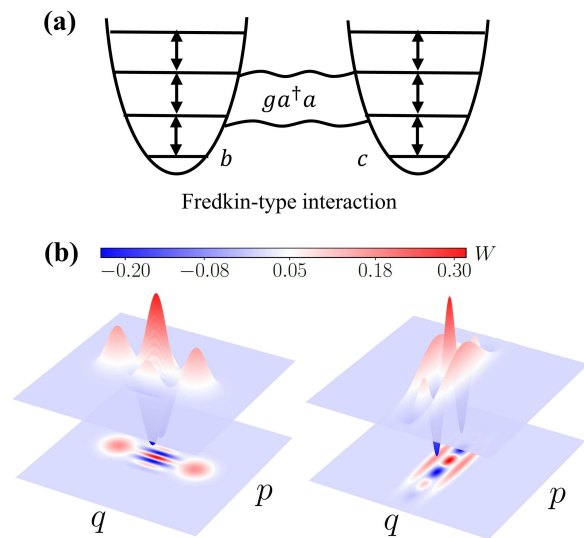}
	\caption{\textbf{(a}): An all-optical platform to prepare the squeezed Schr$\ddot{\mathrm{o}}$dinger-cat states based on the Fredkin-type interaction \cite{GJ1989,RB2016,YY2019} between three optical modes (i.e., $a$,~$b$ and $c$; $g$ is the coupling strength). With the beam-splitter and the cross-Kerr interactions among these three optical modes, the Fredkin-type interaction can be constructed \cite{GJ1989,RB2016,YY2019,XL2022}. \textbf{(b)}: Wigner function of the squeezed Schr$\ddot{\mathrm{o}}$dinger-cat states (here is squeezed even coherent states) in the position-momentum space ($q$,~$p$). Here $q_0=(\alpha+\alpha^*)/\sqrt{2}$ and $p_0=(\alpha-\alpha^*)/i\sqrt{2}$ with the amplitude $\alpha=2$, the squeezing amplitude $r=0$ for the left and $r=1.1$ for the right. The two peaks in the Wigner function correspond to the ``dead cat'' and the ``alive cat'' in the Schr$\ddot{\mathrm{o}}$dinger's paradox of the cat.}
	\label{fig:1}
\end{figure}

As shown in Fig.~\ref{fig:1}(a), we consider an all-optical scheme to prepare squeezed Schr$\ddot{\mathrm{o}}$dinger-cat states based on the Fredkin-type interaction between three optical modes \cite{GJ1989,RB2016,YY2019} represented by annihilation (creation) operators $a$ ($a^{\dagger}$), $b$ ($b^{\dagger}$) and $c$ ($c^{\dagger}$) and their corresponding resonance frequencies $\omega_a$, $\omega_b$ and $\omega_c$. The Fredkin-type interaction describes a two-mode exchange interaction (here modes $b$ and $c$) that depends on the number of photons in another mode (here mode $a$). Based on the Fredkin-type interaction, an all-optical platform to simulate an ultra-strong optomechanical coupling has been proposed \cite{XL2022}. In our scheme, the optical mode $a$ is subject to a coherent two-photon driving and the modes $b$ and $c$ are driven by a coherent driving field, with amplitudes $\Omega_i$, frequencies $\omega_i$ and phases $\phi_i$ ($i=1,~2,~3$). In a rotating frame with respect to $U_{1}=\exp\left\{ i\left[\omega_{1}a^{\dagger}a+\omega_{3}\left(b^{\dagger}b+c^{\dagger}c\right)\right]t\right\}$ , the system Hamiltonian can be written as ($\hbar=1$)
\begin{align}
	H & =\Delta_{a} a^{\dagger} a+\Delta_{b} b^{\dagger} b+\Delta_{c} c^{\dagger} c+g a^{\dagger} a\left(b^{\dagger} c+c^{\dagger} b\right) \notag \\
	&+\Omega_{1}\left(a^{2} e^{-i \phi_{1}}+a^{\dagger 2} e^{i \phi_{1}}\right) \notag \\
	&+\Omega_{2}\left[b e^{i\left(\omega_{2}-\omega_{3}\right) t} e^{-i \phi_{2}}+b^{\dagger} e^{-i\left(\omega_{2}-\omega_{3}\right) t} e^{i \phi_{2}}\right] \notag \\
	&+\Omega_{3}\left(c e^{-i \phi_{3}}+c^{\dagger} e^{i \phi_{3}}\right)
	\label{eq:1}
\end{align}
with detunings $\Delta_{a}=\omega_{a}-\omega_{1}$, $\Delta_{b}=\omega_{b}-\omega_{3}$ and $\Delta_{c}=\omega_{c}-\omega_{3}$. $g$ describes the strength of the Fredkin-type interaction.

In our system, it is assumed that the three optical modes are coupled to the Markovian reservoir. Specifically, the mode $a$ is coupled to a squeezed-vacuum reservoir with a squeezing amplitude $r_e$ and phase $\phi_e$, and the modes $b$ and
$c$ are coupled to an individual thermal reservoir. The squeezed-vacuum reservoir can generally be realized by introducing an auxiliary mode generated by an optical parametric amplification \cite{MOS1997,XY2015}. For an optical mode, the thermal photon number in the thermal reservoir can be ignored safely, then the thermal reservoir can be viewed as a vacuum reservoir. Thus, the evolution of our system in an open environment can be governed by the following master equation, i.e.,
\begin{equation}
\frac{d}{dt}\rho=-i\left[H,\rho\right]+\kappa_a\mathcal{L}(a,\rho)+\kappa_b\mathcal{L}(b,\rho)+\kappa_c\mathcal{L}(c,\rho)
\end{equation}
with the dissipations caused by the system-environment coupling
\begin{align}
	\mathcal{L}(o,\rho) &=\left(N_{e}+1\right) \mathcal{D}[o] \rho+N_{e} \mathcal{D}\left[o^{\dagger}\right] \rho \notag\\
	&-M_{e} \mathcal{G}[o] \rho-M_{e}^{*} \mathcal{G}\left[o^{\dagger}\right] \rho,~o=a,\\
	\mathcal{L}(o,\rho) &=\mathcal{D}[o] \rho,~o=b,c,
\end{align}
where $N_e=\sinh^2(r_e)$ and $M_e=\sinh(r_e)\cosh(r_e)e^{i\phi_e}$ are the mean photon number and the two-photon correlation strength in the squeezing-vacuum reservoir. Here $\mathcal{D}[o]\rho=o\rho o^{\dagger}-(o^{\dagger}o\rho+\rho o^{\dagger}o)/2$, $\mathcal{G}[o]\rho=o\rho o-(oo\rho+\rho oo)/2$, and $\kappa_o$ are the decay rates of the three optical modes.

There is a linear term of the mode $c$ in the Hamiltonian [i.e., $\Omega_{3}(c e^{-i \phi_{3}}+c^{\dagger} e^{i \phi_{3}})$], which can be removed by performing a displacement transformation on the mode $c$ with the displacement operator $D(\eta)=\exp(\eta c^{\dagger}-\eta^* c)$, then the master equation in the displacement representation can be derived as (see Appendix \ref{sec:level1} for details)
\begin{align}
	\frac{d}{d t} \rho_{\mathrm{dis}}&=-i\left[H_{\mathrm{dis}}, \rho_{\mathrm{dis}}\right]\notag\\
	&+\kappa_a\mathcal{L}(a,\rho_{\textrm{dis}})+\kappa_b\mathcal{L}(b,\rho_{\textrm{dis}})+\kappa_c\mathcal{L}(c,\rho_{\textrm{dis}}),
\end{align}
where the density operator in the displacement representation is $\rho_{\mathrm{dis}}=D(\eta)\rho D^{\dagger}(\eta)$ and the Hamiltonian in the displacement representation becomes
\begin{align}
	H_{\mathrm{dis}} &=\Delta_{a} a^{\dagger} a+\Delta_{b} b^{\dagger} b+\Delta_{c} c^{\dagger} c -g a^{\dagger} a\left(b^{\dagger} \eta+b \eta^{*}\right)\notag \\
	&+g a^{\dagger} a\left(b^{\dagger} c+c^{\dagger} b\right)+\Omega_{1}\left(a^{2} e^{-i \phi_{1}}+a^{\dagger 2} e^{i \phi_{1}}\right) \notag\\
	&+\Omega_{2}\left[b e^{i\left(\omega_{2}-\omega_{3}\right) t} e^{-i \phi_{2}}+b^{\dagger} e^{-i\left(\omega_{2}-\omega_{3}\right) t} e^{i \phi_{2}}\right]
	\label{eq:6}
\end{align}
with the amplitude of the displacement operator $\dot{\eta}=-(i\Delta_{c}+\kappa_{c}/2)\eta+i\Omega_{3}e^{i\phi_{3}}$ and its steady value is $\eta_s=\Omega_{3}e^{i\phi_{3}}/(\Delta_{c}-i\kappa_{c}/2)$. Herein we focus on the steady-state region and $\eta_s$ can be a real number by adjusting
the phase $\phi_3$ of the driving field. From Eq.~(\ref{eq:6}), one can find that there is an optomechanical-like interaction $g\eta_sa^{\dagger}a(b^{\dagger}+b)$ with an adjustable coupling $g\eta_s$. By controlling the amplitude $\Omega_3$ of the driving field,
an unltra-strong optomechanical-like interaction can be obtained \cite{XL2022}, which can be equivalent to a Kerr-like interaction and
help to dynamically produce the YSCS \cite{SM1997,SB1997}. However, we focus on preparing other types of (squeezed) Schr$\ddot{\mathrm{o}}$dinger-cat states, which will be shown in the following sections that they have better applications than (squeezed) YSCS.

In our system, the mode $a$ is subject to a coherent two-photon driving, then we can remove the quadratic term in the Hamiltonian of Eq.~(\ref{eq:6}) by carrying out a squeezing transformation on mode $a$ with the squeeze operator $S(\zeta)=\exp[(\zeta a^{\dagger2}-\zeta^{*}a^{2})/2]$. The squeeze parameter $\zeta=r\exp(i\phi_{1})$ with amplitude $r$ and phase $\phi_1$. The master equation in the squeezing representation can be
derived as (see Appendix \ref{sec:level2} for details)
\begin{align}
	\frac{d}{dt}\rho_{\textrm{sq}}&=-i\left[H_{\textrm{sq}},\rho_{\textrm{sq}}\right]\notag\\
	&+\kappa_a\mathcal{L}(a,\rho_{\textrm{sq}})+\kappa_b\mathcal{L}(b,\rho_{\textrm{sq}})+\kappa_c\mathcal{L}(c,\rho_{\textrm{sq}})
	\label{eq:7}
\end{align}
with
\begin{align}
	\mathcal{L}(o,\rho_{\textrm{sq}}) &=\left(N_{\mathrm{eff}}+1\right) \mathcal{D}[o] \rho_{\mathrm{sq}}+N_{\mathrm{eff}} \mathcal{D}\left[o^{\dagger}\right] \rho_{\mathrm{sq}} \notag\\
	&-M_{\mathrm{eff}} \mathcal{G}[o] \rho_{\mathrm{sq}}-M_{\mathrm{eff}}^{*} \mathcal{G}\left[o^{\dagger}\right] \rho_{\mathrm{sq}},~o=a,\\
	\mathcal{L}(o,\rho_{\textrm{sq}}) &=\mathcal{D}[o] \rho_{\mathrm{sq}},~o=b,c,
\end{align}
where the density operator in the squeezing representation is $\rho_{\textrm{sq}}=S(\zeta)D(\eta)\rho D^{\dagger}(\eta)S^{\dagger}(\zeta)$, the mean photon number and the two-photon correlation strength becomes (setting $\phi_1=0$)
\begin{align}
	N_{\mathrm{eff}} &=\sinh ^{2}(r) \cosh ^{2}\left(r_{e}\right)+\cosh ^{2}(r) \sinh ^{2}\left(r_{e}\right) \notag\\
	&+\frac{1}{2} \cos (\phi_e) \sinh (2 r) \sinh \left(2 r_{e}\right), \\
	M_{\mathrm{eff}} &=\left[\cosh \left(r\right) \cosh \left(r_{e}\right)+e^{-i \phi_e} \sinh \left(r\right) \sinh \left(r_{e}\right)\right] \notag\\
	& \times\left[\sinh \left(r\right) \cosh \left(r_{e}\right)+e^{i \phi_e} \cosh \left(r\right) \sinh \left(r_{e}\right)\right],
\end{align}
from which one can see that the mean photon number and the two-photon correlation strength can be suppressed completely (i.e., $N_{\mathrm{eff}},M_{\mathrm{eff}}=0$) when we reasonably adjust the amplitude and phase of the squeezed-vacuum reservoir under the parameter conditions \cite{XY2015}: $r_e=r$ and $\phi_e=\pm n\pi$ ($n=1,3,5...$),  which plays a very important role in suppressing the influence of noise on the system, such as enhanced nonlinear interaction \cite{XY2015}, optical nonreciprocity \cite{LT2022}, state preparation \cite{YH2021,ZC2018,ZC2020,ZC2021}, etc. Besides the Hamiltonian in the squeezing representation can be derived as (dropping
constant terms)
\begin{align}
H_{\mathrm{sq}} &=\omega_{s a} a^{\dagger} a+\Delta_{b} b^{\dagger} b+\Delta_{c} c^{\dagger} c+\frac{g_{p} \eta_{s}}{2}\left(a^{2} b^{\dagger}+a^{\dagger 2} b\right) \notag\\
&+\!\Omega_{2}\!\!\left[\!b e^{i\left(\omega_{2}-\omega_{3}\right) t} e^{-i \phi_{2}}\!+\!b^{\dagger} e^{-i\left(\omega_{2}-\omega_{3}\right) t} e^{i \phi_{2}}\!\right]\! \!+\!\!H_{\mathrm{nr}},\\
H_{\mathrm{nr}} &=\left[\frac{g_{p} \eta_{s}}{2}a^2 b-\frac{g_p}{2}\left(a^{2} b^{\dagger} c+a^{2} c^{\dagger} b\right)\right.\notag\\
&\left.-\left(g_{s}a^{\dagger}a+g\sinh^{2}r\right)\left(b\eta_{s}-bc^{\dagger}\right)\right]+ \mathrm{H.c.},
\end{align}
with the transformed frequency of the mode $a$ $\omega_{sa}=\Delta_a/\cosh(2r)$, the couplings $g_s=g\cosh(2r)$, $g_{p}=g\sinh(2r)$ and the squeeze amplitude $r=\frac{1}{4}\ln\left[\left(\Delta_{a}+2\Omega_{1}\right)/\left(\Delta_{a}-2\Omega_{1}\right)\right]$. The value of the squeeze amplitude $r$ can be adjustable by controlling the amplitude and detuning of the driving field. In the interaction picture with
$U_{2}=\exp\left[i\left(\omega_{sa}a^{\dagger}a+\Delta_{b}b^{\dagger}b+\Delta_{c}c^{\dagger}c\right)t\right]$, the Hamiltonian can be rewritten as
\begin{widetext}
\begin{align}
H_{\mathrm{I}} &=\frac{g_{p} \eta_{s}}{2}\left(a^{2} b^{\dagger}+a^{\dagger 2} b\right)+\Omega_{2}\left(b e^{-i \phi_{2}}+b^{\dagger} e^{i \phi_{2}}\right) +H^{\prime}_{\mathrm{nr}},\label{eq:exa}\\
H^{\prime}_{\mathrm{nr}} &=\left[\frac{g_{p} \eta_{s}}{2}a^2 be^{-2 i \Delta_{b} t}-\frac{g_p}{2}\left(a^{2} b^{\dagger} c e^{-i\Delta_ct}+a^{2} c^{\dagger} b e^{-i(2\Delta_b-\Delta_c)t}\right)\right.\notag\\
&\left.-\left(g_{s}a^{\dagger}a+g\sinh^{2}\!r\right)\left(b\eta_{s}e^{-i\Delta_bt}
-bc^{\dagger}e^{-i(\Delta_b-\Delta_c)t}\right)\right]+ \mathrm{H.c.},
\label{eq:13}
\end{align}
\end{widetext}
where we have assumed that $2\omega_{sa}\!=\!\Delta_{b}$ and $\omega_{2}\!-\!\omega_{3}\!=\!\Delta_{b}$. Under the parameter conditions: $2\Delta_{b}\gg (g_{p}\eta_sn_a\sqrt{n_b}/2$, $2g_{s}\eta_s n_a\sqrt{n_b}$, $2g\eta_s \sinh^2r\sqrt{n_b}$), $\left|\Delta_{b}-\Delta_{c}\right|\gg (g_s n_a\sqrt{n_bn_c}$, $g\sinh^2r\sqrt{n_bn_c}$), and $\left|\Delta_{b}-\Delta_{c}\pm2\omega_{sa}\right|\gg g_pn_a\sqrt{n_bn_c}/2$, with $n_o(o=a,b,c)$ being the dominant excitation numbers in the mode $o$, the Hamiltonian in Eq.~(\ref{eq:13}) becomes the term that oscillates with high frequency and can be safely ignored under the rotating wave approximation (RWA). Then the total Hamiltonian in the interaction picture can be reduced to the following form,
\begin{equation}
H_{\mathrm{I}}=G\left(a^{2}b^{\dagger}+a^{\dagger2}b\right)+\Omega_{2}\left(be^{-i\phi_{2}}+b^{\dagger}e^{i\phi_{2}}\right)
\label{eq:14}
\end{equation}
with the coupling strength $G=g_p\eta_s/2$ between modes $a$ and $b$. From Eq.~(\ref{eq:14}), one can see that it is a Hamiltonian that describes
the degenerate three-wave mixing process of modes $a$ and $b$, where a photon of the mode $b$ is absorbed from (emitted into) the driving field with amplitude $\Omega_2$ and two photons of the mode $a$ are created (annihilated) simultaneously.

From the above discussions, we obtain the degenerate three-wave mixing process from the driven all-optical system with the Fredkin-type interaction under the RWA. To realize the Fredkin-type interaction is the key to the experimental realization of our all-optical scheme for obtaining the degenerate three-wave mixing process. It has been suggested and analyzed that with the beam-splitter and the cross-Kerr interactions among these three optical modes, the Fredkin-type interaction can be obtained and has been realized in experiments \cite{GJ1989,RB2016,YY2019,XL2022}. Besides, the validity of the RWA is manifested in our numerical simulation. As shown in the following sections, we will focus
on the exact Hamiltonian in Eq.~(\ref{eq:exa}) and the approximate Hamiltonian in Eq.~(\ref{eq:14}) for analytical solution and numerical simulation \cite{JR2012,JR2013}.

\section{Deterministic generation of squeezed Schr\"odinger-cat states}\label{sec:3}

At first, we assume that the decay rate of the mode $b$ is large enough (i.e., $\kappa_b\gg\kappa_a$), so that the decay rate of the mode $a$ can be neglected safely. In this case, the mode $b$ can be eliminated adiabatically, then the master equation of the system with the Hamiltonian in Eq.~(\ref{eq:14}) becomes \cite{CC1993,LG1994}
\begin{equation}
\frac{d}{dt}\rho_{a}=-i\left[H_{\textrm{eff}},\rho_{a}\right]+\Gamma_a\mathcal{L}(a^2,\rho_{a}),
\label{eq:15}
\end{equation}
where the effective Hamiltonian is $H_{\textrm{eff}}=iJ(a^{2}-a^{\dagger 2})$ with the effective coupling $J=2\Omega_2 e^{i\phi_2} G/\kappa_b$ describing the rates of the simultaneous decay and excitation of two photons, respectively, $\rho_a$ is the reduced density operator of the mode $a$, and $\Gamma_a=4G^2/\kappa_b$ is the effective two-photon
decay rate. It has been proved analytically that the master equation of Eq.~(\ref{eq:15}) has a steady state for the mode $a$, meanwhile, the steady state only depends on its initial excited number \cite{CC1993,LG1994}. Specifically, for the mode $a$ initially in an even Fock state, its steady state is an ECS with even number distribution, i.e.,
\begin{equation}
\left|\psi_{a}\right\rangle = \mathcal{N}_{e}^{-1/2}\left(\left|\alpha\right\rangle +\left|-\alpha\right\rangle \right)
\end{equation}
with the normalization coefficient $\mathcal{N}_{e}=2\left[1+\exp(-2\left|\alpha\right|^{2})\right]$ and the amplitude $\alpha=\sqrt{-\Omega_{2}e^{i\phi_{2}}/G}$. For the mode $a$ initially in an odd Fock state, it is an OCS with odd number distribution, i.e.,
\begin{equation}
\left|\psi_{a}\right\rangle = \mathcal{N}_{o}^{-1/2}\left(\left|\alpha\right\rangle -\left|-\alpha\right\rangle \right)
\end{equation}
with the normalization coefficient $\mathcal{N}_{o}=2\left[1-\exp(-2\left|\alpha\right|^{2})\right]$. And for a generic initial state of the mode $a$, its steady state will be a mixture of the ECS and the OCS. In an actual experimental system, however, the decay rate of the mode $a$ should be considered. In other words, the lifetime of the generated Schr$\ddot{\mathrm{o}}$dinger-cat states will be limited by the single-photon loss, so the effective two-photon decay rate $\Gamma_a$ should be large to make the mode $a$ can reach its target state with high speed and high fidelity. Unlike other schemes with the two-photon loss, our effective two-photon decay rate $\Gamma_a$ can be significantly enhanced by adjusting the driving fields to obtain a large value of the coupling $G$, which is valuable for an actual system with decays, as shown in the following sections.
\begin{figure*}
\centering\includegraphics[scale=1.1]{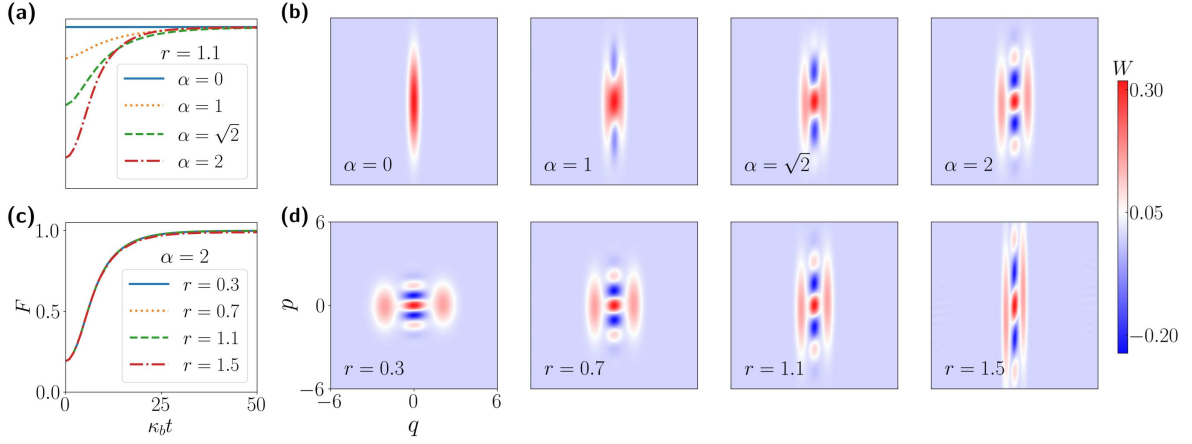}
\caption{\textbf{(a}) and \textbf{(c}): Time evolution of the fidelity $F$ between the actual state $\rho_{\mathrm{app}}$ generated by our system and the exact state $\rho_{\mathrm{exa}}$, with different cat sizes $\alpha$ and squeezed amplitudes $r$. \textbf{(b}) and \textbf{(d}): Plots of Wigner function at steady state with \textbf{(b}) fixed squeezed amplitude $r=1.1$ and different cat sizes $\alpha$, and with \textbf{(d}) fixed cat size $\alpha=2$ and different squeezed amplitudes $r$. In the laboratory framework, the modes $a$, $b$ and $c$ are in squeezed vacuum state, vacuum state and coherent state, respectively. The parameters are set as: $\Delta_a=100\kappa_b,~g=10^{-5}\Delta_a,~\Delta_c=11\Delta_b,~G=0.1\kappa_b,~\kappa_c=\kappa_b$ and $\kappa_a=0$. Other parameters can be gotten from their relations in the text.}
\label{fig:2}
\end{figure*}

In the derivation of the Hamiltonian above, one can find that four unitary transformations (with operators $U_2,~S,~D,$ and $U_1$) have been made, so the state of the mode $a$ in the laboratory framework should be
\begin{align}
	\left|\psi_{a}\right\rangle_{\mathrm{la}}&=U_{1}^{\dagger}D^{\dagger}(\eta)S^{\dagger}(r)U_{2}^{\dagger}\left|\psi_{a}\right\rangle \notag\\
	&=\exp\left(-i\omega_{1}a^{\dagger}at\right)S^{\dagger}(r)\left|\psi_{a}(\alpha^{\prime})\right\rangle,
	\label{eq:18}
\end{align}
where $\left|\psi_{a}(\alpha^{\prime})\right\rangle$ is still a Schr$\ddot{\mathrm{o}}$dinger-cat state with a modified amplitude $\alpha^{\prime}=\alpha \exp(-i\omega_{sa}t)$. Meanwhile, the term $\exp(-i\omega_1a^{\dagger}at)$ only causes the state to rotate in the phase space and does not influence the properties of the state, such as the photon number distribution and the average value of the observable. That is to say, in the laboratory framework, the state of the mode $a$ is a steady squeezed Schr$\ddot{\mathrm{o}}$dinger-cat state in our system. In the following sections, for the sake of convenience, we will consider the state of the mode $a$ in the laboratory framework as $\left|\psi_{a}\right\rangle_{\mathrm{la}}=S^{\dagger}(r)\left|\psi_{a}(\alpha)\right\rangle$ and $\alpha$ as a real number.

\subsection{Squeezed Schr$\ddot{\textbf{o}}$dinger-cat states without the single-photon loss}

To see the quantum features of the generated squeezed Schr$\ddot{\mathrm{o}}$dinger-cat states, we now calculate the Wigner function analytically. The Wigner function is a phase-space quasiprobability distribution \cite{UL1997,MOS1997}, which is defined in the position and momentum space ($q,~p$) as
\begin{equation}
W(q,p)=\frac{1}{2\pi}\int_{-\infty}^{+\infty}dx\left\langle q-\frac{x}{2}\right|\rho_{\mathrm{la}}\left|q+\frac{x}{2}\right\rangle e^{ipx},
\end{equation}
here $\rho_{\mathrm{la}}=\left|\psi_{a}\right\rangle _{\mathrm{la}}\left\langle \psi_{a}\right|$. For the squeezed even coherent state (SECS), the Wigner function is derived as
\begin{equation}
W(q,p)=W_1+W_2+W_{in}
\end{equation}
where
\begin{align}
	W_1&=\frac{\exp\left[-\left(e^{-r}p-p_{0}\right)^{2}-\left(e^{r}q-q_{0}\right)^{2}\right]}{2\pi\left(1+e^{-p_{0}^{2}-q_{0}^{2}}\right)},\label{eq:21}\\
	W_2&=\frac{\exp\left[-\left(e^{-r}p+p_{0}\right)^{2}-\left(e^{r}q+q_{0}\right)^{2}\right]}{2\pi\left(1+e^{-p_{0}^{2}-q_{0}^{2}}\right)},\label{eq:22}\\
	W_{in}&=\frac{\exp\left(-e^{-2r}p^{2}-e^{2r}q^{2}\right)}{\pi\left(1+e^{-p_{0}^{2}-q_{0}^{2}}\right)}\cos\left[2\left(e^{-r}pq_{0}-e^{r}p_{0}q\right)\right], \label{eq:23}
\end{align}
with $q_{0}=\left(\alpha+\alpha^{*}\right)/\sqrt{2}$ and $p_{0}=\left(\alpha-\alpha^{*}\right)/i\sqrt{2}$. Similarly, one can calculate the Wigner function of the squeezed odd coherent state (SOCS) according to its definition. From Eqs.~(\ref{eq:21}-\ref{eq:23}), one can see that the Wigner function exhibits two squeezed peaks at the position-momentum space ($q=\pm e^{-r}q_0$,~$p=\pm e^{r}p_0$). Meanwhile, there are quantum interference and coherence effects between the two peaks, displaying the superposition of both amplitudes and showing rapid oscillations. As shown in Fig.~\ref{fig:1}(b), we plot the Wigner function of the SECS with squeeze amplitudes $r=0$ and $r=1.1$, respectively. Similar to the Schr$\ddot{\mathrm{o}}$dinger’s paradox of the cat, the two peaks in the Wigner function correspond to the ``dead cat'' and the ``alive cat''. One\, can\, see\, that\, the \,Wigner \mbox{functions} show negative values, indicating the nonclassical feature of the squeezed Schr$\ddot{\mathrm{o}}$dinger-cat state.  Moreover, the optical mode $a$ is squeezed on the position $q$  at the expense of expanding in its momentum $p$.

Above we have analyzed the deterministic generation of Schr$\ddot{\mathrm{o}}$dinger-cat states (i.e., squeezed Schr$\ddot{\mathrm{o}}$dinger-cat states in the laboratory framework) with the Hamiltonian in Eq.~(\ref{eq:14}), we now use the exact Hamiltonian in Eq.~(\ref{eq:exa}) to confirm the effectiveness of our all-optical platform to generate the squeezed Schr$\ddot{\mathrm{o}}$dinger-cat states. We define the fidelity $F$ between the approximate state $\rho_{\mathrm{app}}$ generated by our system and the exact state $\rho_{\mathrm{exa}}$ as
\begin{equation}
F=\mathrm{Tr}\left[\sqrt{\sqrt{\rho_{\mathrm{exa}}}\rho_{\mathrm{app}}\sqrt{\rho_{\mathrm{exa}}}}\right],
\end{equation}
here $\rho_{\mathrm{exa}}=\left|\psi_{a}\right\rangle _{\mathrm{la}}\left\langle \psi_{a}\right|$. When the fidelity $F$ approaches $1$, it means that the state generated by our system is the same as the ideal state.  As shown in Fig.~\ref{fig:2}, we plot the time evolution of the fidelity $F$ with different cat sizes $\alpha$ and squeezed amplitudes $r$, as well as the corresponding Wigner functions at steady state. One can see that the optical mode $a$, as expected, is steered into a stable squeezed cat state with high fidelity, meanwhile, the cat size and the squeezed amplitude have a good adjustability by controlling the driving fields in our system.

\begin{figure}[H]
	\centering\includegraphics[scale=0.29]{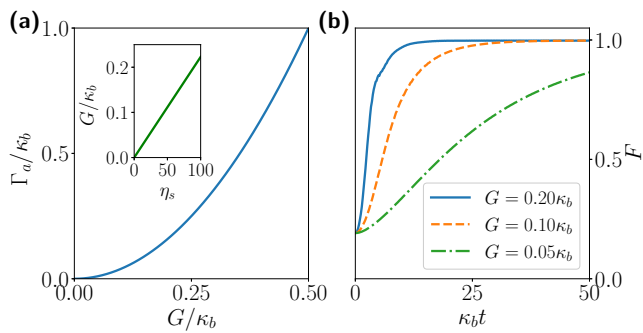}
	\caption{\textbf{(a}): Effective two-photon decay rate $\Gamma_a$ is plotted as a function of coupling $G$, and the inset shows the change of the coupling $G$ with the parameter $\eta_s$ of the driving field. \textbf{(b}): Evolution of the fidelity $F$ with cat size $\alpha=2$, squeeze amplitude $r=1.1$, and different couplings $G~(0.20\kappa_b,~0.10\kappa_b,~0.05\kappa_b)$. Other parameters are the same as those in Fig.~\ref{fig:2}.}
	\label{fig:3}
\end{figure}

Meanwhile, our system has the ability to enhance the coupling $G$ by adjusting the driving fields so that an increased two-photon loss $\Gamma_a$ can be obtained, which helps to reduce the time required to reach the stable squeezed Schr$\ddot{\mathrm{o}}$dinger-cat states, as shown in Fig.~\ref{fig:3}. From the curves, one can clearly find that with a larger two-photon loss, the optical mode $a$ can reach the steady squeezed Schr$\ddot{\mathrm{o}}$dinger-cat state with higher speed. Otherwise it will take a longer time to evolve, which is undoubtedly detrimental to practical applications. Thus, with our
all-optical system, parameter adjustable steady squeezed Schr$\ddot{\mathrm{o}}$dinger-cat states with high fidelity and speed can be obtained, which will have application value in quantum information, quantum metrology and so on.

\subsection{Squeezed Schr$\ddot{\textbf{o}}$dinger-cat states with the single-photon loss}

So far, we have not considered the decay rate (i.e., the single-photon loss) of the optical mode $a$. The decay rate of the optical mode $a$ will limit the lifetime of the generated squeezed Schr$\ddot{\mathrm{o}}$dinger-cat states. In this section, we analyze the evolution of the squeezed Schr$\ddot{\mathrm{o}}$dinger-cat states under the single-photon loss.

\begin{figure*}
\centering\includegraphics[scale=0.9]{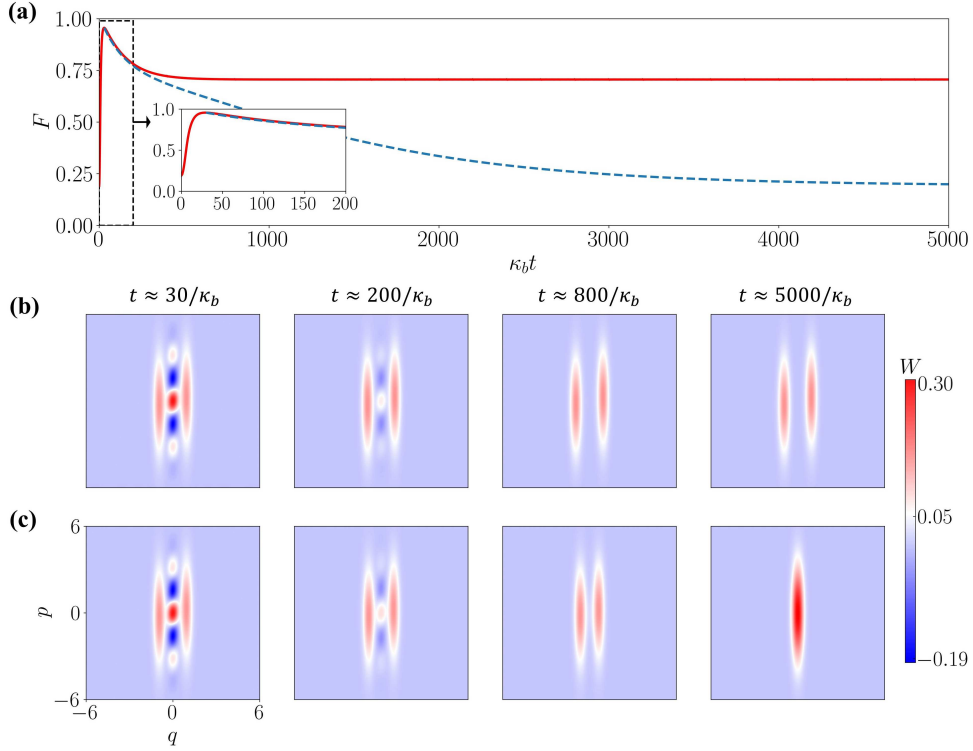}
\caption{\textbf{(a}): Time evolution of the fidelity $F$ in the system with simultaneous single-photon and two-photon losses of the mode $a$, i.e., solid cure, and with only the single-photon loss of the mode $a$, i.e., dashed cure. \textbf{(b}) and \textbf{(c}): Plots of Wigner function of the mode $a$ at different times ($t\approx 30/\kappa_b,\, 200/\kappa_b,\, 800/\kappa_b,\, 5000/\kappa_b)$, corresponding to the solid and dashed cures in \textbf{(a}), respectively. Other parameters are the same as those in Fig.~\ref{fig:2} except for $\alpha=2$, $r=1.1$ and $\kappa_a=10^{-3}\kappa_b$.}
\label{fig:4}
\end{figure*}

From Eq.~(\ref{eq:15}), one can find that in our system with a large decay rate of the mode $b$, the mode $b$ has been eliminated adiabatically, so the system only has the mode $a$ that subjects to the two-photon loss and evolves to a squeezed Schr$\ddot{\mathrm{o}}$dinger-cat state in the laboratory framework. In practice, the mode $a$ also subjects to a single-photon loss, that is, the master equation in Eq.~(\ref{eq:15}) should be changed as
\begin{equation}
\frac{d}{dt}\rho_{a}=-i\left[H_{\textrm{eff}},\rho_{a}\right]+\Gamma_a\mathcal{L}(a^2,\rho_{a})+\kappa_a\mathcal{L}(a,\rho_{a}),
\label{eq:26}
\end{equation}
which causes the the mode $a$ to evolve to a squeezed Schr$\ddot{\mathrm{o}}$dinger-cat state with certain distortion. Now we use the
exact Hamiltonian [i.e., Eq.~(\ref{eq:exa})] to simulate the evolution of the mode $a$ under its single-photon loss. As shown in Fig.~\ref{fig:4}(a) (see the solid cure), we plot the time evolution of the fidelity $F$ in the system under the single-photon loss of the mode $a$ [corresponding to simultaneous single-photon and two-photon loss in Eq.~(\ref{eq:26})]. From the curve, one can see that in an open system with a modest single-photon loss, the mode $a$ can be prepared to the squeezed Schr$\ddot{\mathrm{o}}$dinger-cat state with a good fidelity ($F\approx 0.96$). Then the fidelity decreases gradually and stays at a stable value in the end. Meanwhile, we plot the Wigner function of the mode $a$ at different times as shown in Fig.~~\ref{fig:4}(b). One can clearly see that with the time evolution, the quantum interference and coherence effects between the two squeezed peaks disappear gradually. Moreover, the negative values of the Wigner function also disappear, which are in sharp contrast to Fig.~\ref{fig:2}. At last, the two squeezed peaks remain in the plot of Wigner function. Actually,
due to the simultaneous existence of single-photon and two-photon losses of the mode $a$ [see Eq.~(\ref{eq:26})], one can find that it will be in a mixture of (approximately) two squeezed coherent states in the end after decaying out the coherence.

Besides, we also analyze the evolution of the mode $a$ with the two-photon loss turned off [i.e., $G=0$ by adjusting the driving field, see Fig.~\ref{fig:3}(a)] when it evolves to a squeezed Schr$\ddot{\mathrm{o}}$dinger-cat state with a good fidelity. In this case, the mode $a$ is only subjected to the single-photon loss. Under a single-photon loss, the density operator (here SECS) of the mode $a$ will decay during the evolution, and can be written as \cite{SH2006}
\begin{widetext}
\begin{align}
	\rho_{\mathrm{doc}}	&=N_e^{-1}S^{\dagger}\left\{ \left|\alpha e^{-\kappa_{a}t/2}\right\rangle \left\langle \alpha e^{-\kappa_{a}t/2}\right|+\left|-\alpha e^{-\kappa_{a}t/2}\right\rangle \left\langle -\alpha e^{-\kappa_{a}t/2}\right|\right. \notag \\
	&+\left.\exp\left[-2\alpha^{2}\left(1-e^{-\kappa_{a}t}\right)\right]\left[\left|\alpha e^{-\kappa_{a}t/2}\right\rangle \left\langle -\alpha e^{-\kappa_{a}t/2}\right|+\left|-\alpha e^{-\kappa_{a}t/2}\right\rangle \left\langle \alpha e^{-\kappa_{a}t/2}\right|\right]\right\}S,
\end{align}
\end{widetext}
where the decay of the diagonal term is accompanied by the energy loss, while the decay of the non-diagonal term is accompanied by the attenuation of coherence (i.e., decoherence). After a long time evolution with a single-photon loss, the coherence of the generated squeezed Schr$\ddot{\mathrm{o}}$dinger-cat states and the energy of the system will all disappear, so that the mode $a$ will evolve into a squeezed vacuum state in the end. The lifetime of the squeezed Schr$\ddot{\mathrm{o}}$dinger-cat state can be approximated as $\tau=1/2\left|\alpha\right|^2\kappa_a$ according to the above equation \cite{SH2006}. As shown in Fig.~\ref{fig:4}(a) (see the dashed curve), we plot the evolution of the fidelity $F$ in the system with only the single-photon loss of the mode $a$ when it is evolved to the squeezed Schr$\ddot{\mathrm{o}}$dinger-cat state with a good fidelity. From the curve, one can see that the value of fidelity decreases monotonously and approaches its initial value, that is, the mode $a$ is evolved to a squeezed vacuum state. Moreover, we also plot the Wigner function at the same times as Fig.~\ref{fig:4}(b), as shown in Fig.~\ref{fig:4}(c). One can see that the difference between them is that
the two squeezed peaks of the Wigner function decay into one squeezed peak in the end.

In the experiment, in order to quickly prepare squeezed Schr$\ddot{\mathrm{o}}$dinger-cat states and significantly prolong their lifetimes, we must enhance the effective two-photon loss $\Gamma_a$ of the mode $a$ and choose a small decay rate $\kappa_a$ of the mode $a$. For a modest decay rate of the optical mode \mbox{$\kappa_a\sim 10^{5}\,\mathrm{Hz}$} \cite{MA2014}, the lifetime of the state will be about the order of microseconds. However, with a smaller decay rate, the lifetime of the state will be further increased. For example, with an extreme cavity decay rate ($\sim 0.2\,\mathrm{Hz}$), Schr$\ddot{\mathrm{o}}$dinger-cat states with the order of milliseconds have been prepared in experiment \cite{SD2008}. Besides, due to the extremely weak spin relaxation, long-lived atomic Schr$\ddot{\mathrm{o}}$dinger-cat states with about $3$ seconds have also been predicted in theory by engineering the two-atom decay \cite{WQ2021}. That is to say, reducing the corresponding decay rate as much as possible will be the key to generate the quantum states in experiment.

\section{Phase estimation with squeezed Schr\"odinger-cat states}\label{sec:4}

Above we have shown that deterministic squeezed Schr$\ddot{\mathrm{o}}$dinger-cat states (i.e., SECS, SOCS and their mixture) with high fidelity can be prepared in our all-optical platform by engineering the two-photon loss. The light field in a squeezed Schr$\ddot{\mathrm{o}}$dinger-cat state can be applied to a variety of quantum technologies, such as quantum computation, quantum metrology,
etc. As an example, in this section, we exploit the generated squeezed Schr$\ddot{\mathrm{o}}$dinger-cat states to estimate the phase in the optical interferometer and compare the QFI with those using the perfect squeezed Schr$\ddot{\mathrm{o}}$dinger-cat states and states without squeezing.

\begin{figure}[b]
    \centering\includegraphics[scale=0.45]{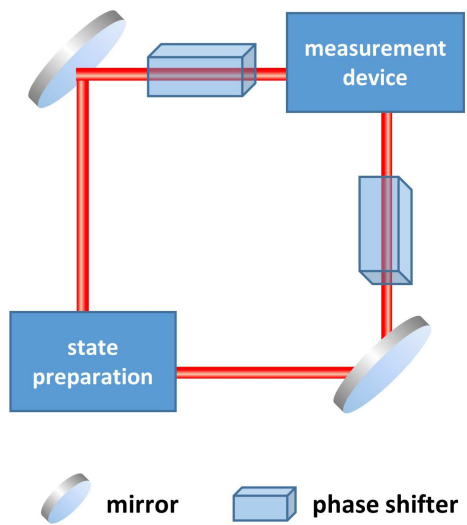}
	\caption{Optical interferometer model used for the estimation of the phase. A prepared quantum state is sent into the interferometer, which contains an unknown relative phase ($\varphi_1-\varphi_2$) generated by two linear phase shifters between its two arms. The phase information will be imprinted on the initial quantum state and be measured at the output ports.}
	\label{fig:5}
\end{figure}
As shown in Fig.~\ref{fig:5}, there are two linear phase shifters that contain an unknown relative phase ($\varphi_1-\varphi_2$) between the arms of the interferometer. The phase information will be imprinted on the initial quantum state (i.e., squeezed Schr$\ddot{\mathrm{o}}$dinger-cat states) in the input of the interferometer by the unitary evolution, i.e.,
\begin{equation}
\left|\Psi\right\rangle =U\left|\Psi_{0}\right\rangle ,
\end{equation}
where the unitary operator $U=\exp\left[i(\varphi_{1}a_{1}^{\dagger}a_{1}+\varphi_{2}a_{2}^{\dagger}a_{2})\right]$ with $a_{1,2}\,(a^{\dagger}_{1,2})$ representing annihilation (creation) operators in the arms 1 and 2. $\left|\Psi_{0}\right\rangle =\left|\psi_{1}\right\rangle \otimes\left|\psi_{2}\right\rangle$ is the input state of the interferometer, with $\left|\psi_{1,2}\right\rangle$ being the squeezed Schr$\ddot{\mathrm{o}}$dinger-cat states [see Eq.~(\ref{eq:18})]. Here we focus on the relative phase between the two arms, then for a pure and path-symmetric initial state, the evolution operator $U$ can be reduced to \cite{MJ2012} $U=\exp\left[i\phi_{-}O_{-}\right]$ with $\varphi_{-}=\varphi_{1}-\varphi_{2}$ and $O_{-}=(a_{1}^{\dagger}a_{1}-a_{2}^{\dagger}a_{2})/2$. For the unitary evolution of the initial state $\left|\Psi_{0}\right\rangle$, the QFI can be defined as \cite{JL2019,JL2022,JL2015}
\begin{equation}
\mathcal{F}=4\left\langle \Psi_{0}\left|\Delta^{2}\mathcal{H}\right|\Psi_{0}\right\rangle,
\end{equation}
where $\mathcal{H}=i(\partial_{\phi_{-}}U^{\dagger})U$ and $\Delta^{2}\mathcal{H}=(\mathcal{H}-\left\langle \mathcal{H}\right\rangle )^{2}$. With the expression of the operator $U$, one can calculate the QFI as
\begin{equation}
\mathcal{F}=2\left[\mathrm{Var_{\Psi_{0}}}(a_{i}^{\dagger}a_{i})-\mathrm{Cov_{\Psi_{0}}}(a_{1}^{\dagger}a_{1},a_{2}^{\dagger}a_{2})\right],\;i=1~\textrm{or}~2,
\label{eq:29}
\end{equation}
where $\mathrm{Var_{\Psi_{0}}}(\bullet)$ and $\mathrm{Cov_{\Psi_{0}}}(\bullet)$ are the variance and the covariance in the state $\left|\Psi_{0}\right\rangle$, respectively. From Eq.~(\ref{eq:29}), one can find that the value of the QFI only depends on the properties of the initial input state. As the initial state considered here is separable, so we have $\mathrm{Cov_{\Psi_{0}}}(a_{1}^{\dagger}a_{1},a_{2}^{\dagger}a_{2})=0$, and the QFI is reduced to the variance of the photon number operator $a_i^{\dagger}a_i$.

Based on Eq.~(\ref{eq:29}), we calculate the the expression of the QFI analytically. Specifically, for the SECS, the QFI and the total average photon number $N$ of the two arms are
\begin{align}
	\mathcal{F}&=\sinh^{2}(2r)-2\alpha^{2}\mathrm{sech}(\alpha^{2})\sinh(4r-\alpha^{2}) \notag\\
	&+2\alpha^{4}\cosh^{2}(2r)\mathrm{sech}^{2}(\alpha^{2}),\\
	N&=2\left[\sinh^{2}(r)-\alpha^{2}\mathrm{sech(\alpha^{2})\sinh(2r-\alpha^{2})}\right].
\end{align}
If the squeeze amplitude $r=0$ (i.e., ECS), they are reduced to $\mathcal{F}=2\alpha^{4}\mathrm{sech^{2}}(\alpha^{2})+N$ and $N=2\alpha^{2}\tanh(\alpha^{2})$, respectively. One can find that even if in the low-photon-number regime the QFI can beat the value of the standard quantum limit (SQL, i.e., $\mathcal{F}\sim N$), but its scaling can only reach the SQL in the limit of the large photon number. For the SOCS, one has
\begin{align}
	\mathcal{F}&=\sinh^{2}(2r)+2\alpha^{2}\mathrm{csch}(\alpha^{2})\cosh(4r-\alpha^{2}) \notag\\
	&-2\alpha^{4}\cosh^{2}(2r)\mathrm{csch}^{2}(\alpha^{2}),\\
	N&=2\left[\sinh^{2}(r)+\alpha^{2}\mathrm{csch}(\alpha^{2})\mathrm{\cosh(2r-\alpha^{2})}\right].
\end{align}
If the squeeze amplitude $r=0$ (i.e., OCS), they are reduced to $\mathcal{F}=-2\alpha^{4}\mathrm{csch^{2}}(\alpha^{2})+N$ and $N=2\alpha^{2}\coth^{2}(\alpha^{2})$, respectively.  One can also find that in the limit of the large photon number the scaling of the QFI can reach the SQL, but it is inferior to the SQL in the low-photon-number regime. For comparison, we also calculate the case of the squeezed Yurke-Stoler coherent state (SYSCS), i.e, $\left|\psi\right\rangle =S^{\dagger}(r)\left[\left(\frac{1+i}{2}\right)\left|\text{\ensuremath{\alpha}}\right\rangle +\left(\frac{1-i}{2}\right)\left|\text{\ensuremath{-\alpha}}\right\rangle \right]$. The QFI and the average photon number become
\begin{align}
	\mathcal{F}&=\sinh^{2}(2r)+2e^{-4r}\alpha^{2},\\
	N&=2\left[\sinh^{2}(r)+\alpha^{2}e^{-2r}\right].
\end{align}
Similarly, when the squeeze amplitude $r=0$ (i.e., YSCS), they are reduced to $\mathcal{F}=N$ and $N=2\alpha^{2}$, respectively. One can clearly find that with the YSCS, the scaling of the QFI is only the SQL.

\begin{figure}[t]
	\centering\includegraphics[scale=0.16]{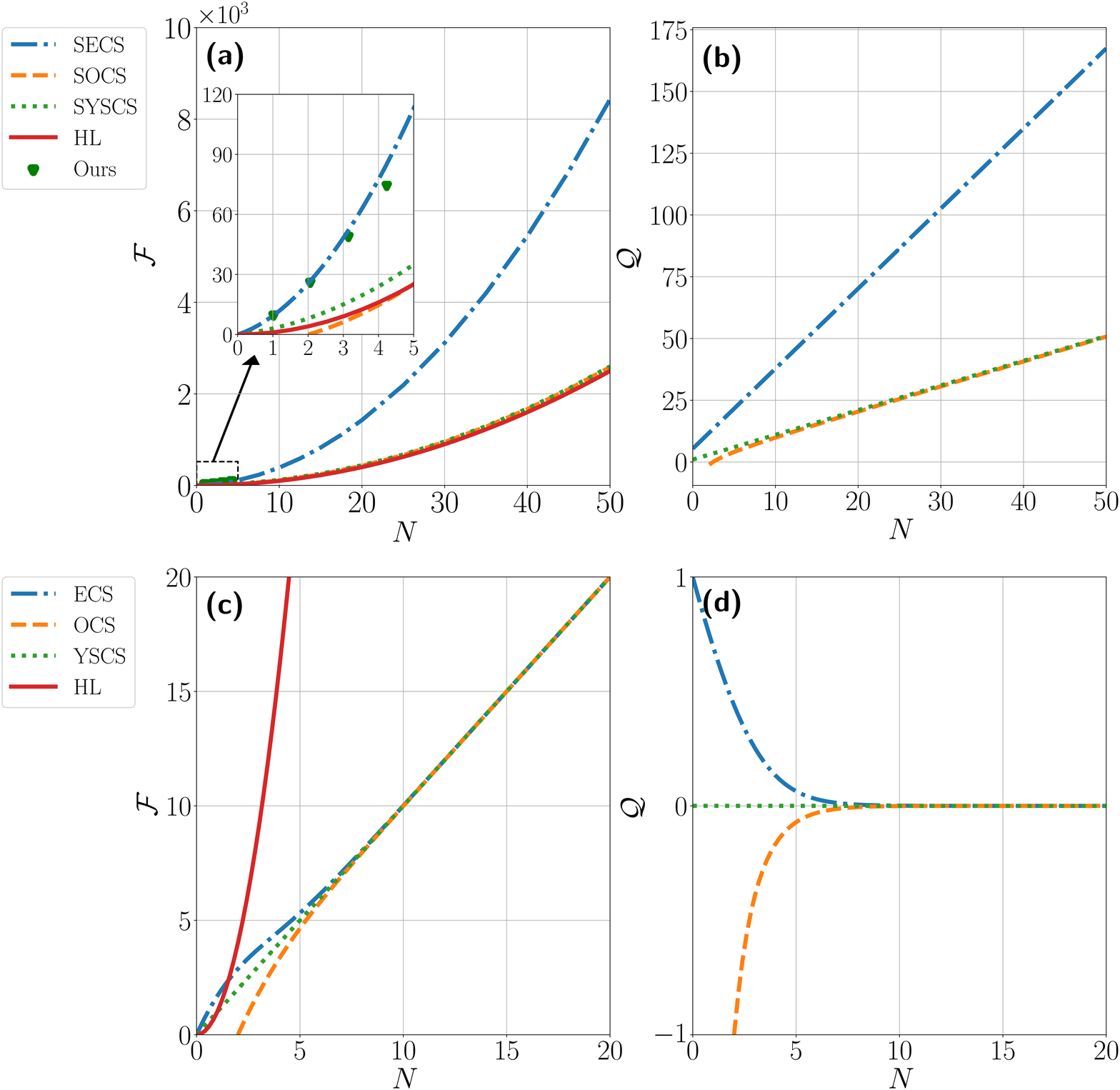}
	\caption{\textbf{(a}) and \textbf{(c}): Quantum Fisher information (QFI) $\mathcal{F}$ is plotted as a function of the average photon number $N$ with different initial states [i.e., \textbf{(a}) squeezed even coherent state (SECS), squeezed odd coherent state (SOCS) and squeezed Yurke-Stoler coherent state (SYSCS); \textbf{(c}) even coherent state (ECS), odd coherent state (OCS) and Yurke-Stoler coherent state (YSCS)]. \textbf{(b}) and \textbf{(d}): Mandel parameter $\mathcal{Q}$ of the Schr$\ddot{\mathrm{o}}$dinger-cat states with and without squeezing is plotted as a function of the average photon number $N$. HL represents the Heisenberg limit. Inverted triangle symbol represents the QFI with the SECS generated by our all-optical system. Please note that the average photon number of the SOCS and the OCS is greater than or equal to 2 (i.e., $N\geq2$) due to their photon number distributions.}
	\label{fig:6}
\end{figure}

From the expressions of the QFI using the squeezed Schr$\ddot{\mathrm{o}}$dinger-cat states, one can see that compared with the one without squeezing, there are two parameters, i.e., $r$ and $\alpha$, can be chosen to increase the value of the QFI. To see the scaling of these QFIs, one can optimize the two parameters to maximize the
QFI for a fixed average photon number. As shown in Fig.~\ref{fig:6}(a), we plot the QFIs using the above three squeezed Schr$\ddot{\mathrm{o}}$dinger-cat states by numerically optimizing $r$ and $\alpha$ with the each fixed average photon number. From the curves, one can see that the value of the QFI using the SECS
is better than those using the SOCS and the SYSCS. Specifically, in the case of the large photon number, the values of the QFIs using these three states are all higher than the value of the HL (corresponding $N^2$), especially the SECS. In the case of the low photon number, however, the SOCS is lower than the HL [see the inset of Fig.~~\ref{fig:6}(a)]. To see the scaling of the QFIs using these three types of squeezed Schr$\ddot{\mathrm{o}}$dinger-cat states in the limit of a large photon number, we numerically fit the relation of the QFI with the average photon number. For the SECS,
\begin{equation}
\mathcal{F}\approx6.42N + 3.24N^2;
\label{eq:37}
\end{equation}
for the SOCS,
\begin{equation}
\mathcal{F}\approx-4.40-3.01N^{\frac{1}{2}} + 2.28 N + N^2;
\label{eq:38}
\end{equation}
and for the SYSCS,
\begin{equation}
\mathcal{F}\approx2N + N^2.
\label{eq:39}
\end{equation}
From the above equations, one can find that in the limit of a large photon number, the scaling of the QFI with these states is all the HL. Moreover, in the case of a low photon number, the QFI using the SECS has significant advantages to estimate the phase of the optical interferometer due to the factor improvement over the HL, which is very important for fragile systems that cannot withstand large photon fluxes, such as spin ensembles \cite{FW2013}, atoms \cite{MK2008}, molecules \cite{MP2011}, biological systems \cite{MA2013}, and so on. Besides, in Fig.~\ref{fig:6}(c), we also plot the QFI for the ECS, the OCS and the YSCS with different average photon numbers. Compared with Fig.~\ref{fig:6}(a), the scaling of the QFI with these states is only the SQL in the limit of a large number of photons. Thus, one can see the importance of squeezing the Schr$\ddot{\mathrm{o}}$dinger-cat states.

One can understand this squeezing-enhanced phase estimation as follows. As pointed out in Ref.~\cite{JS2015}, the QFI of Eq.~(\ref{eq:29}) can be rewritten as
\begin{equation}
\mathcal{F}=N(1+\mathcal{Q})(1-\mathcal{J}),
\end{equation}
where the Mandel parameter of the mode $a_i$ has the form $\mathcal{Q}=\left[\mathrm{Var_{\Psi_{0}}}(a_{i}^{\dagger}a_{i})-\left\langle a_{i}^{\dagger}a_{i}\right\rangle\right]/\left\langle a_{i}^{\dagger}a_{i}\right\rangle $, and $\mathcal{J}=\mathrm{Cov_{\Psi_{0}}}(a_{1}^{\dagger}a_{1},a_{2}^{\dagger}a_{2})/\mathrm{Var_{\Psi_{0}}}(a_{i}^{\dagger}a_{i})$ ranges from -1 to 1 and is zero for the separable states considered here. So the Mandel parameter for the pure and separable initial states can be written as
\begin{equation}
\mathcal{Q}=\frac{\mathcal{F}}{N}-1,
\end{equation}
then we can get the expressions of the above squeezed Schr$\ddot{\mathrm{o}}$dinger-cat states according to the fitting formulas in Eqs.~(\ref{eq:37}-\ref{eq:39}). Specifically, for the SECS,
\begin{equation}
\mathcal{Q}\approx5.42 + 3.24N;
\end{equation}
for the SOCS
\begin{equation}
\mathcal{Q}\approx -4.40N^{-1} -3.01N^{-\frac{1}{2}}+ 1.28 + N;
\end{equation}
and for the SYSCS
\begin{equation}
\mathcal{Q}\approx 1 + N.
\end{equation}
Besides, one can also get the versions without squeezing. As shown in Figs.~\ref{fig:6}(b) and (d), the Mandel parameter of the Schr$\ddot{\mathrm{o}}$dinger-cat states with and without squeezing are plotted as a function of the average photon number. One can find that the Mandel parameters of the squeezed Schr$\ddot{\mathrm{o}}$dinger-cat states are proportional to the average photon number, except for the SOCS in the low photon number [see Fig.~\ref{fig:6}(b)], which can contribute to bringing their QFI values to the HL in the limit of a large photon number. However, for the Mandel parameter of the Schr$\ddot{\mathrm{o}}$dinger-cat states without squeezing [see Fig.~\ref{fig:6}(d)], one can clearly find that with the increase of the average photon number, the Mandel parameters all go to zero, whereas the YSCS remains zero. This causes their QFI values only to approach the SQL, which is in stark contrast to those of the corresponding squeezing Schr$\ddot{\mathrm{o}}$dinger-cat states. At last,
we compare the QFI of the squeezed Schr$\ddot{\mathrm{o}}$dinger-cat state (here SECS) generated by our all-optical platform with the QFI of the above perfect one in the low-photon-number regime, as shown in the inset of Fig.~\ref{fig:6}(a). One can see that due to the high fidelity of the generated squeezed Schr$\ddot{\mathrm{o}}$dinger-cat state in our system, the QFI value is in good agreement with the perfect one. Meanwhile, it can have an
order of magnitude factor improvement over the value of the HL in the low-photon-number regime.

\section{Conclusions}\label{sec:5}

In conclusion, we propose an all-optical platform based on the Fredkin-type interaction to deterministically generate the squeezed Schr$\ddot{\mathrm{o}}$dinger-cat states with high speed and high fidelity. Meanwhile, we get the analytical expression of the Wigner function of the squeezed Schr$\ddot{\mathrm{o}}$dinger-cat state, and show the evolution of the Wigner function in an open system in detail. Besides, we exploit the squeezed Schr$\ddot{\mathrm{o}}$dinger-cat states generated by our system to estimate the phase in an optical interferometer, and find
that in the limit of a large photon number the QFI can all reach the HL. Especially, the QFI with the SECS can have an order of magnitude factor improvement over the HL in the low-photon-number regime, which is important for fragile systems that cannot withstand large photon fluxes.

\begin{acknowledgments}
We thank Dr.~Chang-Sheng Hu, Dr.~Ye-Hong Chen and Dr.~Wei Qin for technical support and helpful discussions. This work was supported by the National Natural Science Foundation of China (Grants No. 11935012, No. 11875231,
No. 11805073, No. 12175075, and No. 11947069), the National Key Research and Development Program of China (Grants No. 2017YFA0304202 and No. 2017YFA0205700), and the Scientific Research Fund of Hunan Provincial Education Department (Grant No. 20C0495).
\end{acknowledgments}

\appendix

\section{\label{sec:level1} Master equation in the displacement representation}

To remove the linear term of the mode $c$ [i.e., $\Omega_{3}(c e^{-i \phi_{3}}+c^{\dagger} e^{i \phi_{3}})$] of Eq.~(\ref{eq:1}), we perform a displacement transformation on the mode $c$, i.e., $D(\eta)\rho D^{\dagger}(\eta)=\rho_{\mathrm{dis}}$, with the displacement operator $D(\eta)=\exp(\eta c^{\dagger}-\eta^{*}c)$. Then we have
\begin{align}
	\dot{\rho}&=\frac{d}{d t}\left[D^{\dagger}(\eta) \rho_{\mathrm{dis}} D(\eta)\right] \notag \\
	&=\dot{D^{\dagger}}(\eta)\rho_{\mathrm{dis}}D(\eta)
	+D^{\dagger}(\eta)\dot{\rho}_{\mathrm{dis}}D(\eta)+D^{\dagger}(\eta)\rho_{\mathrm{dis}}\dot{D}(\eta)\notag \\
	&=-i\left[H,D^{\dagger}(\eta) \rho_{\mathrm{dis}} D(\eta)\right]+\kappa_a\mathcal{L}(a,D^{\dagger}(\eta) \rho_{\mathrm{dis}} D(\eta))\notag \\
	&+\!\!\kappa_b\mathcal{L}(b,D^{\dagger}(\eta) \rho_{\mathrm{dis}} D(\eta))\!\!+\!\!\kappa_c\mathcal{L}(c,D^{\dagger}(\eta) \rho_{\mathrm{dis}} D(\eta)),
\end{align}
so
\begin{align}
	D^{\dagger}(\eta)\dot{\rho}_{\mathrm{dis}}D(\eta)&=-\left[\dot{D^{\dagger}}(\eta)\rho_{\mathrm{dis}}D(\eta)
	+D^{\dagger}(\eta)\rho_{\mathrm{dis}}\dot{D}(\eta)\right] \notag\\
	&-i\left[H,D^{\dagger}(\eta) \rho_{\mathrm{dis}} D(\eta)\right]\notag\\
	&+\kappa_a\mathcal{L}(a,D^{\dagger}(\eta) \rho_{\mathrm{dis}} D(\eta))\notag \\
	&+\kappa_b\mathcal{L}(b,D^{\dagger}(\eta) \rho_{\mathrm{dis}} D(\eta))\notag\\
	&+\kappa_c\mathcal{L}(c,D^{\dagger}(\eta) \rho_{\mathrm{dis}} D(\eta)).
	\label{eq:a2}
\end{align}
After some tedious calculations, one can get the master equation in the displacement representation [i.e., Eq.~(\ref{eq:6})],
\begin{align}
	\frac{d}{d t} \rho_{\mathrm{dis}}&=-i\left[H_{\mathrm{dis}}, \rho_{\mathrm{dis}}\right]\notag\\
	&+\kappa_a\mathcal{L}(a,\rho_{\textrm{dis}})+\kappa_b\mathcal{L}(b,\rho_{\textrm{dis}})+\kappa_c\mathcal{L}(c,\rho_{\textrm{dis}}),
\end{align}
by multiplying $D(\eta)$ on the left and $D^{\dagger}(\eta)$ on the right of the Eq.~(\ref{eq:a2}) and eliminating the linear terms in the transformed Liouvillian. This process leads to the following equation for the amplitude $\eta$,
\begin{equation}
\dot{\eta}=-(i\Delta_{c}+\kappa_{c}/2)\eta+i\Omega_{3}e^{i\phi_{3}},
\end{equation}
then its steady value is $\eta_s=\Omega_{3}e^{i\phi_{3}}/(\Delta_{c}-i\kappa_{c}/2)$.

\section{\label{sec:level2} Master equation in the squeezed representation}
To remove the quadratic term of the mode $a$ [i.e., $\Omega_1(a^2 e^{-i\phi_1}+a^{\dagger 2}e^{i\phi_1})$] of Eq.~(\ref{eq:6}), we perform a squeezing transformation on the mode $a$, i.e., $S(\zeta)\rho_{\textrm{dis}}S^{\dagger}(\zeta)=\rho_{\textrm{sq}}$, with the squeeze operator
$S(\zeta)=\exp[(\zeta a^{\dagger2}-\zeta^{*}a^{2})/2]$. Similar to the above displacement transformation, one can get the master equation in the squeezed representation [i.e., Eq.~(\ref{eq:7})],
\begin{align}
	\frac{d}{dt}\rho_{\textrm{sq}}&=-i\left[H_{\textrm{sq}},\rho_{\textrm{sq}}\right]\notag\\
	&+\kappa_a\mathcal{L}(a,\rho_{\textrm{sq}})+\kappa_b\mathcal{L}(b,\rho_{\textrm{sq}})+\kappa_c\mathcal{L}(c,\rho_{\textrm{sq}})
\end{align}
by eliminating the non-diagonal term of the transformed Hamiltonian, which leads to the following equation for the amplitude $r$,
\begin{equation}
\Delta_{a}\sinh\left(2r\right)/2-\cosh(2r)\Omega_1=0,
\end{equation}
then we have $r=\frac{1}{4}\ln\left[\left(\Delta_{a}+2\Omega_{1}\right)/\left(\Delta_{a}-2\Omega_{1}\right)\right]$.

\end{document}